\documentclass[12pt,preprint,notoc,nohyper]{QCEMDA} % 10pt is ignored!

\usepackage{epsfig,multicol,bbm}

\newcommand\fverb{\setbox\pippobox=\hbox\bgroup\verb}
\newcommand\fverbdo{\egroup\medskip\noindent%
            \fbox{\unhbox\pippobox}\ }
\newcommand\fverbit{\egroup\item[\fbox{\unhbox\pippobox}]}
\newbox\pippobox
\title{Emission of scalar particles \\ from cylindrical black holes}

\author{H. Gohar and K. Saifullah  \\

Department of Mathematics, Quaid-i-Azam University, Islamabad,
Pakistan \\

Electronic address: \email{saifullah@qau.edu.pk}}

\preprint{}  % OR: \preprint{Aaaa/Mm/Yy\\Aaa-aa/Nnnnnn}
\abstract{We study quantum tunneling of scalar particles from black
strings. For this purpose we apply WKB approximation and
Hamilton-Jacobi method to solve the Klein-Gordon equation for
outgoing trajectories. We find the tunneling probability of outgoing
charged and uncharged scalars from the event horizon of black
strings, and hence the Hawking temperature for these black
configurations.}

\begin{document}

\section{Introduction}

Black holes are objects in this universe with such a strong
gravitational field that even light cannot escape from them. The
important breakthrough in the field of black hole physics occurred
when Stephen Hawking showed that quantum mechanically black holes
emit radiations \cite{1, 2}. Due to the strong gravitational field
and vacuum fluctuations at the event horizon of the black hole,
virtual particles-anti particles are created. Here we can have three
types of scenarios: (a) both the particles fall into the hole, (b)
both of them escape from the event horizon, and (c) one particle
falls into the hole while the other escapes. The particle that
escapes appears as the Hawking radiation. The negative energy
particle that falls into the black hole reduces the mass, charge and
the angular momentum of the black hole. As a result, the black hole
shrinks. This particle must go into the black hole to conserve
energy.

After Hawking's discovery these thermal radiations have been studied
for different black bodies. There are different methods to derive
Hawking radiation and Hawking temperature. These can be studied, for
example, by calculating the Bogoliubov transformation \cite{1, new}
between the initial and final states of ingoing and outgoing
radiation. The Wick rotation method \cite{new2, new22} is also used
for investigating Hawking radiation. Recently, the black hole
tunneling method (\cite{3}- \cite{15}), anomaly method \cite{new3}
and the technique of dimensional reduction \cite{Um} have been used
to investigate Hawking radiation and Hawking temperature. The
radiation spectrum from black holes contains all types of particles
including scalar particles \cite{Ya,13}. Here, we have used the
black hole tunneling method to derive the tunneling probability of
scalar particles from black strings (\cite{16}- \cite{19}). In order
to do this we solve the Klein-Gordon equation by using WKB
approximation and complex path integration. As a result we obtain
Hawking temperature also.

\section{Black strings}

The Einstein field equations have a large number of solutions. Here,
we discuss some special solutions of these equations, which are
exact with negative cosmological constant, called black strings or
cylindrical black holes.

A four dimensional metric with $g_{\mu \nu } \,\,\, (\mu ,\nu
=0,1,2,3)$ is given by \cite{18}
\begin{equation}
ds^{2}=g_{\mu \nu }dx^{\mu }dx^{\nu }=g_{mn}dx^{m}dx^{n}+e^{-4\phi
}dz^{2}, \label{qqw}
\end{equation}%
where $g_{mn}$ and $\phi $ are metric functions, $m,n=0,1,2$ ,
$x^{\mu }=\left( t,r,\theta ,z\right) $ and $z$ is the Killing
coordinate. We will write a cylindrically symmetric metric by taking
the $\theta $ coordinate also in Killing direction from Eq.
(\ref{qqw}). We consider the Einstein-Hilbert action in four
dimensions with a negative cosmological constant in the
presence of an electromagnetic field. The total action is given by%
\begin{equation}
S+S_{em}=\frac{1}{16\pi G}\int d^{4}x\sqrt{-g}\left( R-2\Lambda \right) -%
\frac{1}{16\pi }\int d^{4}x\sqrt{-g}F^{\mu \nu }F_{\mu \nu }.
\label{1a}
\end{equation}
Here, $S$ is the Einstein-Hilbert action in four dimensions,
$S_{em}$ is the action for electromagnetic field, $R$ is the Ricci
scalar, $g$ the determinant of the metric tensor, $\Lambda $ the
cosmological constant, $G$ the gravitational constant, and the
Maxwell tensor $F_{\mu \nu }$ is given by
\begin{equation}
F_{\mu \nu }=\partial _{\mu }A_{\nu }-\partial _{\nu }A_{\mu },
\label{2a}
\end{equation}%
where $A_{\nu }$ is vector potential and is given by $A_{\nu
}=-h(r)\delta _{\nu }^{0},$ $h(r)$ being an arbitrary function of
the radial coordinate $r$ . Here, we take the solution of the
Einstein-Maxwell equations with cylindrical symmetry. The line
element for static charged black string with negative cosmological
constant in the presence of electromagnetic field becomes
\cite{18,19}

\begin{equation}
ds^{2}=-(\alpha ^{2}r^{2}-\frac{b}{\alpha r}+\frac{c^{2}}{\alpha ^{2}r^{2}}%
)dt^{2}+(\alpha ^{2}r^{2}-\frac{b}{\alpha r}+\frac{c^{2}}{\alpha ^{2}r^{2}}%
)^{-1}dr^{2}+r^{2}d\theta ^{2}+\alpha ^{2}r^{2}dz^{2},  \label{3a}
\end{equation}
where
\begin{eqnarray}
\alpha ^{2} &=&\frac{-1}{3}\Lambda ,  \label{5a} \\
b &=&4GM,  \label{6a} \\
c^{2} &=&4GQ^{2},  \label{7a} \\
h(r) &=&\frac{2Q}{\alpha r}+\textrm{const.,}  \label{8a} \\
-\infty  &< &t < \infty ,0\leq r < \infty ,-\infty < z < \infty
,0\leq \theta \leq 2\pi .
\end{eqnarray}
Here, $Q$ is the linear charge density per unit length of the $z$
line and $M$ is mass per unit length of the z line of black string.
The event (outer) horizon can be found by putting $\alpha ^{2}r^{2}-\frac{b}{\alpha r}+\frac{%
c^{2}}{\alpha ^{2}r^{2}}=0$ and is given by \cite{18}
\begin{equation}
r=r_{+}=\frac{b^{\frac{1}{3}}\sqrt{s}+\sqrt{2\sqrt{s^{2}-4p^{2}-s}}}{2\alpha
},  \label{9a}
\end{equation}%
where
\begin{eqnarray*}
s &=&\left(
\frac{1}{2}+\frac{1}{2}\sqrt{1-4(\frac{4p^{2}}{3})^{3}}\right) ^{
\frac{1}{3}}+\left(
\frac{1}{2}-\frac{1}{2}\sqrt{1-4(\frac{4p^{2}}{3})^{3}}
\right) ^{\frac{1}{3}}, \\
p^{2} &=&\frac{c^{2}}{b^{\frac{4}{3}}}.
\end{eqnarray*}%
If we put $Q=0$ in Eq. (\ref{3a}) we obtain the simplest case of
black strings, which contains only one parameter, which is the mass
of black string. The line element for this case is given by
\begin{equation}
ds^{2}=-(\alpha ^{2}r^{2}-\frac{b}{\alpha r})dt^{2}+(\alpha
^{2}r^{2}-\frac{b }{\alpha r})^{-1}dr^{2}+r^{2}d\theta ^{2}+\alpha
^{2}r^{2}dz^{2}.
\end{equation}
The outer horizon for this black string can be found by putting
$\alpha ^{2}r^{2}-\frac{b}{\alpha r}=0,$ which gives
\begin{equation}
r=r_{+}=\frac{b^{\frac{1}{3}}}{\alpha }.  \label{11aa}
\end{equation}

\section{Quantum tunneling of scalar particles from black strings}

In order to work out the tunneling probability of scalar particles
from the event horizon of black strings we use the Klein-Gordon
equation for a scalar field $\Psi $ given by

\begin{equation}
g^{\mu \upsilon }\partial _{\mu }\partial _{\upsilon }\Psi -\frac{m^{2}}{%
\hslash ^{2}}\Psi =0.  \label{2}
\end{equation}
Using WKB approximation, we assume an ansatz of the form

\begin{equation}
\Psi (t,r,\theta ,z)=e^{\left( \frac{i}{\hslash }I(t,r,\theta
,z)+I_{1}(t,r,\theta ,z)+O(\hslash )\right) },  \label{3}
\end{equation}
for Eq. (\ref{2}) which can be written as
\begin{equation}
g^{00}\partial _{t}\partial _{t}\Psi +g^{11}\partial _{r}\partial
_{r}\Psi +g^{22}\partial _{\theta }\partial _{\theta }\Psi
+g^{33}\partial _{z}\partial _{z}\Psi -\frac{m^{2}}{\hslash
^{2}}\Psi =0.  \label{4}
\end{equation}%
Now by using Eq. (\ref{3}) in Eq. (\ref{4}) and evaluating term by
term\ in the highest order of $\hslash $ and dividing by the
exponential term and
multiplying by $\hslash ^{2},$ we get%
\begin{equation}
0=-(\alpha ^{2}r^{2}-\frac{b}{\alpha r})^{-1}(\partial
_{t}I)^{2}+(\alpha ^{2}r^{2}-\frac{b}{\alpha r})(\partial
_{r}I)^{2}+\frac{1}{r^{2}}(\partial _{\theta }I)^{2}+\frac{1}{\alpha
^{2}r^{2}}(\partial _{z}I)^{2}.  \label{9}
\end{equation}%
Keeping in view the Killing fields, $\partial _{t},$ $\partial
_{\theta }$ and $\partial _{z},$ of the background spacetime, we
separate the variables and consider a solution for Eq. (\ref{9}) of
the form
\begin{equation}
I(t,r,\theta ,z)=-Et+W(r)+J_{1}\theta +J_{2}z+K,  \label{10}
\end{equation}%
where $E,$ $J_{1}$, $J_{2}$ and $K$ are constants. Here, we are only
considering the radial trajectories. Using Eq. (\ref{10}) in Eq.
(\ref{9}) yields after simplification
\begin{equation}
W^{\prime }(r)=\pm \sqrt{-\frac{g^{00}}{g^{11}}\left( E^{2}+\frac{g^{22}}{%
g^{00}}(J_{1})^{2}+\frac{g^{33}}{g^{00}}(J_{2})^{2}+\frac{1}{g^{00}}%
m^{2}\right) }.  \label{18}
\end{equation}%
Integrating this and substituting the values of $g^{\mu \nu }$ gives,%
\begin{equation}
W(r)=\pm \int \frac{\sqrt{E^{2}-f(r)\left( m^{2}+\frac{(J_{1})^{2}+\frac{%
(J_{2})^{2}}{\alpha ^{2}}}{r^{2}}\right) }}{f(r)}dr,  \label{19}
\end{equation}
where
\begin{equation}
f(r)=\alpha ^{2}r^{2}-\frac{b}{\alpha r}.
\end{equation}
We have to integrate Eq. (\ref{19}) around the pole at the event
horizon, $r_{+}=b^{\frac{1}{3}}/\alpha ,$ We use the residue theory
for semi circle yielding
\begin{equation}
W_{\pm }(r)=\pm \frac{\pi i}{f^{\prime
}(r_{+})}\sqrt{E^{2}-f(r_{+})\left(
m^{2}+\frac{(J_{1})^{2}+\frac{(J_{2})^{2}}{\alpha
^{2}}}{r_{+}^{2}}\right) }. \label{23}
\end{equation}%
As $f(r_{+})=0,$ the above equation reduces to%
\begin{equation}
W_{\pm }(r)=\pm \frac{\pi iE}{f^{\prime }(r_{+})},  \label{24}
\end{equation}%
which implies that
\begin{equation}
Im   W_{\pm }(r)=\pm \frac{\pi E}{f^{\prime }(r_{+})}, \label{25}
\end{equation}
where
\begin{equation}
f^{\prime }(r_{+})=2\alpha ^{2}r_{+}+\frac{b}{\alpha r_{+}^{2}}=3\alpha b^{%
\frac{1}{3}}.  \label{26}
\end{equation}%
The probabilities of crossing the horizon from inside to outside and
outside to inside are given by \cite{5, 6}
\begin{eqnarray}
P_{emission} &\varpropto &\exp \left( \frac{-2}{\hbar }Im   I\right)
=\exp \left( \frac{-2}{\hbar }(Im   W_{+}+Im   K)\right) ,
\label{27} \\
P_{absorption} &\varpropto &\exp \left( \frac{-2}{\hbar }Im I\right)
=\exp \left( \frac{-2}{\hbar }(Im   W_{-}+Im   K)\right) .
\label{28}
\end{eqnarray}
We know that the probability of any incoming particles crossing the
horizons and entering the black hole is one, so it is necessary to
set
\begin{equation}
Im   K=-Im   W_{-}  \label{29} ,
\end{equation}%
in the above equations. From Eq. (\ref{24}), we have%
\begin{equation}
W_{+}=-W_{-}.  \label{30}
\end{equation}%
This means that the probability of a particle tunneling from inside
to outside the horizon is
\begin{equation}
\Gamma =\exp \left( \frac{-4}{\hbar }Im   W_{+}\right) . \label{31}
\end{equation}%
Thus, by using Eq. (\ref{25}), (by choosing $\hbar =1),$ we get
\begin{equation}
\Gamma =\exp \left( -\frac{4\pi E}{f^{\prime }(r_{+})}%
\right) .  \label{34}
\end{equation}%
This is the probability of the outgoing scalar particle from the
event horizon, $r=r_{+}.$ We can find the Hawking temperature, by
comparing Eq. (\ref{34}) with the Boltzmann factor  \cite{5, 6},
$\Gamma =\exp (-\beta E),$ where $E$ is the energy of the particle
and $\beta $ is the inverse of Hawking temperature. Thus we get
\begin{equation}
T_{H}=\frac{f^{\prime }(r_{+})}{4\pi },  \label{35}
\end{equation}%
or%
\begin{equation}
T_{H}=\frac{3\alpha b^{\frac{1}{3}}}{4\pi }.  \label{new3}
\end{equation}

\section{Quantum tunneling of scalar particles from charged black strings}

To study the quantum tunneling from charged black strings (Eq.
(\ref{3a}))
for scalar field $\Psi ,$ we use the charged Klein-Gordon equation%
\begin{equation}
\frac{1}{\sqrt{-g}}\left( \partial _{\mu }-\frac{iq}{\hbar }A_{\mu
}\right) \left( \sqrt{-g}g^{\mu \upsilon }(\partial _{\nu
}-\frac{iq}{\hbar }A_{\nu })\Psi \right) -\frac{m^{2}}{\hslash
^{2}}\Psi =0.  \label{42}
\end{equation}%
Proceeding as before, we apply WKB approximation and assume the
field of the form given in Eq. (\ref{3}). Substituting this in Eq.
(\ref{42}) and keeping terms only in the leading order of $\hbar $
and dividing by exponential term and multiplying by $\hslash ^{2}$
gives
\begin{equation}
g^{tt}(\partial _{t}I+qh(r))^{2}+g^{rr}(\partial
_{r}I)^{2}+g^{\theta \theta }(\partial _{\theta
}I)^{2}+g^{zz}(\partial _{z}I)^{2}+m^{2}=0.  \label{59a}
\end{equation}%
Again assuming a solution of the type of Eq. (\ref{10}) for the
above equation and solving for $W(r)$ we get
\begin{equation}
W_{\pm }(r)=\pm \int \frac{\sqrt{(-E+qh(r))^{2}-f(r)\left(
m^{2}+\frac{ (J_{1})^{2}+ (J_{2})^{2}/ \alpha ^{2}} {r^{2}}\right)
}}{f(r)}dr, \label{70}
\end{equation}
where%
\begin{equation}
f(r)=\alpha ^{2}r^{2}-\frac{b}{\alpha r}+\frac{c^{2}}{\alpha
^{2}r^{2}}.
\end{equation}
Using the complex integration techniques, the integral around the
simple
pole at the event horizon given by Eq. (\ref{9a}) yields%
\begin{equation}
W_{\pm }(r)=\pm \frac{\pi i(-E+qh(r_{+}))}{f^{\prime }(r_{+})}.
\label{77}
\end{equation}
This implies that
\begin{equation}
Im   W_{\pm }(r)=\pm \frac{\pi (-E+qh(r_{+}))}{f^{\prime }(r_{+})},
\label{78}
\end{equation}
where
\begin{eqnarray}
f^{\prime }(r_{+}) &=&2\alpha ^{2}r_{+}+\frac{b}{\alpha r_{+}^{2}}-\frac{%
2c^{2}}{\alpha ^{2}r_{+}^{3}},  \label{79} \\
h(r_{+}) &=&\frac{2Q}{\alpha r_{+}}.  \label{80}
\end{eqnarray}
Thus the tunneling probability of scalar particles from the charged
black string comes out to be
\begin{equation}
\Gamma =\exp \left( -\frac{4\pi (-E+qh(r_{+}))}{\hbar f^{\prime
}(r_{+})}\right) .  \label{86}
\end{equation}%
From this, we can find the Hawking temperature by comparing this
with the Boltzmann factor of particle energy
\begin{equation}
T_{H}=\frac{f^{\prime }(r_{+})}{4\pi },  \label{87}
\end{equation}%
where $f^{\prime }(r_{+})$ is given in Eq. (\ref{79}). So%
\begin{equation}
T_{H}=\frac{1}{4\pi }\left( 2\alpha ^{2}r_{+}+\frac{b}{\alpha r_{+}^{2}}-%
\frac{2c^{2}}{\alpha ^{2}r_{+}^{3}}\right) ,  \label{vvv}
\end{equation}%
which is consistent with the literature \cite{15, 20}.

\section{Conclusion}

In this paper we have studied Hawking radiation of scalar particles
from uncharged and charged black strings. By using Hamilton-Jacobi
method we have solved the charged and uncharged Klein-Gordon
equations. In order to do this we have employed WKB approximation to
Klein-Gordon equation to derive the tunneling probability of
outgoing particles. At the end, by comparing with the Boltzmann
factor of energy for the particles, we have derived the Hawking
temperature for these black configurations. These results are found
to be consistent with the literature. If we put $Q=0$ in Eq.
(\ref{vvv}), the temperature reduces to that for the uncharged case
in Eq. (\ref{new3}).

%\acknowledgments

%We are thankful to Douglas Singleton for his comments.

\end{document}